\documentclass[aps,preprint]{revtex4}%
\usepackage{amsfonts}
\usepackage{amsmath}
\usepackage{amssymb}
\usepackage{graphicx}%
\providecommand{\U}[1]{\protect\rule{.1in}{.1in}}

\begin{document}
\preprint{HEP/123-qed}
\title[Short title for running header]{Solar neutrinos, helicity effects and new affine gravity with torsion}
\author{Diego Julio Cirilo-Lombardo}
\affiliation{Bogoliubov Laboratory of Theoretical Physics, Joint Institute for Nuclear
Research 141980, Dubna(Moscow Region), Russian Federation}
\pacs{PACS number}

\begin{abstract}
New $f(R,T)$ model of gravitation, introduced previously by the author, is
considered. \ It is based on an affine geometrical construction in which the
torsion is a dynamical field, the coupling is minimal and the theory is
Lorentz invariant by construction. It was shown that the Dirac equation
emerges from the same space time and acquires a modification (coupling-like)
of the form $\gamma^{\alpha}j\frac{1-d}{d}\gamma_{5}h_{\alpha}$, with
$h_{\alpha}$ the torsion axial vector, $j$ a parameter of pure geometrical
nature and $d$, the spacetime dimension. In the present work it is shown that
this interaction produces a mechanism of spin (helicity) flipping, with its
consequent weak symmetry violation. The cross section of this process is
explicitly calculated and a logaritmical energy dependence (even at high
energies) is found. This behavior is reminiscent of similar computations made
by Hans Bethe in the context of neutrino astrophysics. These results are
applied to the solar neutrino case and compared with similar results coming
from a gravitational model with torsion of string theory type and within the
standard model context respectively.

\end{abstract}
\email{diego777jcl@gmail.com, diego@theor.jinr.ru; Tel: +74 9621 66938/Fax: +74 9621 65084}
\volumeyear{year}
\volumenumber{number}
\issuenumber{number}
\eid{identifier}
\date[Date text]{date}
\received[Received text]{date}

\revised[Revised text]{date}

\accepted[Accepted text]{date}

\published[Published text]{date}

\startpage{101}
\endpage{102}
\maketitle
\tableofcontents

\section{ Introduction}

\label{sec1} As is commonly suggested, solutions based on neutrino
\textquotedblleft spin flip\textquotedblright\ in the Sun's magnetic fields
are proposed to explain the observed solar neutrino deficit. Dependence of the
survival probability on energy and significant regeneration effect (day/night
asymmetry) are not observed in solar neutrino detectors. In the last ten
years, an increasing interest in the neutrino physics is shown by the
scientific community. Four (probably related) aspects involving neutrinos are
constantly the target of investigations: the solar neutrino problem , the CP
and CPT violations effects and the anomalous momentum [3]. From the
astrophysical point of view neutrino has a non-zero magnetic moment, the
neutrino helicity can be flipped when it passes through a region with magnetic
field perpendicular to the direction of propagation. It means that the
left-handed neutrino that is active in SM would change into a right-handed one

In this paper, we attack mostly the first one, proposing an explanation for
the solar neutrino problem, alternative to the proposals based on more
standard approaches given, for example, in [16] and to the string theoretical
arguments, involving a torsion potential, given for example in [18]. This
motivation is justified by the very important analysis of the problem given by
reference [13] where the bounds of the axial parameters are actualized and new
experiments in this direction are suggested; and reference [14](by the same
group) where bounds to the non-standard interactions of the muonic neutrinos
and quarks are given.

The arguments on behalf of our explanation arise in a new model of gravitation
based on a pure affine geometrical construction. The geometrical Lagrangian of
the theory contains dynamically as main geometrical object, $\ $a generalized
curvature $\mathcal{R=}\det(\mathcal{R}_{\ \mu}^{a})$ (coming from a higher
dimensional group manifold generally based in SU(2,2N)) and is defined as
follows
\begin{equation}
L_{g}=\sqrt{det\mathcal{R}_{\ \mu}^{a}\mathcal{R}_{a\nu}}=\sqrt{detG_{\mu\nu}}
\tag{1}%
\end{equation}
where after the breaking of the symmetry and due to the reductivity of the
geometry, the generalized curvature becomes to
\begin{equation}
\mathcal{R}_{\ \mu}^{a}=\lambda\left(  e_{\ \mu}^{a}+f_{\ \mu}^{a}\right)
+R_{\ \mu}^{a}\qquad\left(  M_{\mu}^{a}\equiv e^{a\nu}M_{\nu\mu}\right)
\tag{2}%
\end{equation}
and the Lagrangian $L_{g}$ coming to
\begin{equation}
\sqrt{Det\ \mathcal{R}_{\ \mu}^{a}\mathcal{R}_{a\nu}}=\sqrt{Det\left[
\lambda^{2}\left(  g_{\mu\nu}+f_{\ \mu}^{a}f_{a\nu}\right)  +2\lambda
R_{\left(  \mu\nu\right)  }+2\lambda f_{\ \mu}^{a}R_{[a\nu]}+R_{\ \mu}%
^{a}R_{a\nu}\right]  }, \tag{3}%
\end{equation}
where $R_{\mu\nu}=R_{(\mu\nu)}+R_{[\mu\nu]}$ and $f_{\ \mu}^{a}$ , in a sharp
contrast with the tetrad field $e_{\ \mu}^{a}$, carries the symmetry $e_{a\mu
}f_{\ \nu}^{a}=f_{\mu\nu}=-f_{\nu\mu}.$-- see [5,6,7] for more mathematical
and geometrical details of the theory. As we have been previously pointed out,
the original definition given by us of the Lagrangian as a measure containing
$\mathcal{R}_{\ \mu}^{a}$ is reminiscent to have the same essence as in the
pioneering unified field theories, in particular the Eddington one.

As was shown in [5,6,7], in this specific model of $f\left(  R,T\right)
$gravity the Dirac equation is derived from the same spacetime manifold, and
acquires a modification (coupling-like) of the form
\begin{equation}
\gamma^{\alpha}j\frac{1-d}{d}\gamma_{5}h_{\alpha}, \tag{4}%
\end{equation}
with $h_{\alpha}$ the torsion axial vector, $j$ a parameter of pure
geometrical nature and $d$, the spacetime dimension. Here the torsion is a
dynamical field, and the theory is Lorentz invariant by construction.

In the following, we will show that this interaction produces a mechanism of
spin flipping and we will calculate explicitly the cross section for this
process. This cross section, in sharp contrast with the string and standard
model cases[18,16], depends logarithmically on the energy, even at high
energies, which is a reminiscent of similar computations made by Hans Bethe
time ago [2] considering astrophysical neutrinos.

The behavior of the cross section energy dependence is very important when
considering solar neutrinos detected experimentally because the number of
events naturally depends on the energy threshold. For example, the Cl detector
of the Davis group has a threshold of 0.8 MeV, with a lower count (28\%), than
the KAMIOKA\ neutrino detector (with a threshold of 7.5 MeV), and the SAGE and
GALLEX (with threshold of 0.23 MeV) have higher counts (51-62\%) [12,19,17,1].
The importance of this research is more that evident due the well known
following reasons. With the increasing level of precision of \ the neutrino
experiments, in high energy and in astrophysics the determination of neutrino
parameters is the crucial point. Today, the main focus from the
phenomenological point of view, is devoted to the determination with high
precision \ of the oscillation parameters (as the testing of non-oscillation
effects and possible subleading oscillations), non standard neutrino
interactions (NSI) and as spin-flavour conversions. Although magnetic field
effects, that are the first candidates to produce density fluctuations into
the radiative zone of the sun, doesn't not modifies the robustness of the
quantitative analysis of neutrino oscillations.

In our manuscript we focus on the case of neutrinos endowed with non-standard
interactions making evident due the presence of the torsion as dynamical field
. These are also a natural outcome of many neutrino mass models and can be in
general of two types: flavour changing (FC) and non-universal (NU). Models of
Seesaw type are the best examples models leading structures of the lepton
mixing matrix that are non trivial ones. These matrix characterize the charged
and neutral current weak interactions. Then, the NSI\ that are induced by
gauge transformations can, even with massless neutrinos, CP\ and leptonic
flavor. Models where the masses of the neutrinos coming from radiative
corrections as in superunified models the NSI\ surely appears.

The organization of this paper is as follows: in Section \ref{sec2} we obtain,
by geometrical methods, the solution of a particle at rest in Minkowski
spacetime (where the curvature effects are not important but torsion certainly
exists). Considerations about the form of the geometrical Lagrangian of the
theory and the nature (electric or magnetic) of the axial vector torsion are
given with some technical details. In Section \ref{sec3}, also making some
exhaustive use of geometrical manipulations, we explicitly compute the cross
section corresponding to a neutrino-hadron interaction in the presence of
torsion. Section \ref{sec4} is devoted to the application of the formulation
of the previous section to the computation of probability of spin flipping for
the solar neutrinos, and comparing these results with similar results coming
from a gravitational model with torsion of string theory type, and with
totally antisymmetric torsion. Finally, Sections \ref{sec5} and \ref{sec6} are
devoted to summarize the obtained results.

\section{Statement of the problem}

\label{sec2}

Consider the approximation in which, avoiding strong curvature effects, we
deal only with torsion fields. Seeking for spherical symmetric solutions, the
line element under consideration will be, for instance,
\begin{equation}
ds^{2}=-dt^{2}+dr^{2}+r^{2}\left(  d\theta^{2}+\sin^{2}\theta\ d\varphi
^{2}\right)  \tag{5}%
\end{equation}

Therefore, the tensor $G_{\mu\nu}=-2\lambda\left(  g_{\mu\nu}+f_{\mu}%
^{a}f_{a\nu}\right)  $ of the geometrical Lagrangian (1) -- stated in [5,6,7]
-- reduces to $\left.  G_{\mu\nu}\right\vert _{Mink.}=-2\lambda\left(
\eta_{\mu\nu}+f_{\mu}^{a}f_{a\nu}\right)  $. To solve the equation of our
problem we use the Cartan's structure method with the natural orthonormal
system: $\omega^{0}=dt,\quad\omega^{1}=dr,\quad\omega^{2}=rd\theta
,\ \omega^{3}=r\sin\theta d\varphi,$ as usual for spherical symmetry. We use
the Palatini principle that simultaneously determines the connection required
for the spacetime symmetry and the dynamical field equations. The specific
form of the action S (or the Lagrangian as we have been commented in the
Introduction) is arbitrary but from this action $S$ necessarily we must reach
the G-invariant conditions, namely, the intersection of the 4 dimensional
Lorentz group $L_{4},$ the symplectic $Sp\left(  4\right)  $ and the almost
complex group $K\left(  4\right)  )$without prior assumption. Then, the
Einstein, Dirac and Maxwell equations need to arise from $S$ as a causally
connected closed system. \ Is very important to regard here that the
antisymmetric $f$ field forming part of the geometrical construction can be
associated to the physical electromagnetic field , namely $\varphi_{\mu\nu}$
by means the following relation%

\begin{equation}
f_{\mu\nu}\equiv\frac{1}{2}\varepsilon_{\mu\nu\rho\sigma}\overline{\varphi
}^{\rho\sigma}=\ast\overline{\varphi}_{\mu\nu}\tag{6}%
\end{equation}
where $\overline{\varphi}^{\mu\nu}$ is the inverse tensor to $\overline
{\varphi}_{\mu\nu}$ and $\ast$is the Hodge operator defined into the four
dimensional spacetime. It is important to note that $f_{\mu\nu}$ is
adimensional, corresponding to our choice for $g_{\mu\nu}$ in the Lagrangian
of the theory. That means that, physically, there exists an "absolute field",
namely $j$, that fulfills (analog to the $b$ field in the Born-Infeld theory )
the double role of homogenizing the units and put a maximum limit to the
magnitude of the fields: $\frac{\varphi_{\mu\nu}}{j}\equiv$ $\overline
{\varphi}_{\mu\nu}$, this is the meaning of the bar over the physical fields.
This field $j$ should play a fundamental physical role in the phenomenology of
the spin flipping, as we will see when computing the cross section of the
neutrino-hadron interaction. We will work, for simplicity, with $f_{\mu\nu}$
reminding that the physical electromagnetic fields $\varphi_{\mu\nu}$ are the
respective dual due (6): for electric $f_{\mu\nu}$ corresponds physical
magnetic $\varphi_{\mu\nu}$and viceversa.

The axial vector $h_{\mu}$, dual of the total antisymmetric torsion field
$T_{\mu\nu\varrho}$, plays a fundamental role in the Dirac equation derived in
[5,6,7] from the same spacetime manifold $M$. It modifies the anomalous
momentum $g$ of the elementary fermionic particles and, due to the symmetries
of the problem, we assume $h=h\left(  r,\theta\right)  $ (geometrical
notation). The dynamical equations of the $f$ field are comparable in form to
the Born-Infeld case [9,10]. Then, as a natural ansatz, $f$ will be assumed to
have the form%
\begin{equation}
f=f_{31}\omega^{3}\wedge\omega^{1}+f_{23}\omega^{2}\wedge\omega^{3}%
+f_{10}\omega^{1}\wedge\omega^{0}+f_{20}\omega^{2}\wedge\omega^{0} \tag{7}%
\end{equation}
and $\mathbb{F\equiv}\frac{\mathbb{\partial L}_{G}}{\partial f}.$

The dynamical equation and the Bianchi identities are given, respectively, by
$d\ast\mathbb{F}=0$ and $df=0$. Explicitly we have
\begin{align}
\partial_{\theta}\left(  f_{31}r\sin\theta\right)  +\partial_{r}\left(
f_{23}r^{2}\sin\theta\right)   &  =0\tag{8}\\
\partial_{r}\left(  f_{20}r\right)  -\partial_{\theta}f_{10}  &  =0\nonumber\\
& \nonumber\\
\partial_{\theta}\left(  \mathbb{F}_{20}r\sin\theta\right)  +\partial
_{r}\left(  \mathbb{F}_{10}r^{2}\sin\theta\right)   &  =0\tag{9}\\
\partial_{r}\left(  \mathbb{F}_{31}r\right)  -\partial_{\theta}\mathbb{F}%
_{23}  &  =0\nonumber
\end{align}
The $G_{\mu\nu}$ tensor can be easily performed taking (in Cartesian
coordinates) the following form
\begin{equation}
G_{\mu\nu}=\left(
\begin{array}
[c]{cccc}%
-1+f_{10}^{2}+f_{20}^{2} & 0 & 0 & f_{20}f_{23}+f_{13}f_{10}\\
0 & 1+f_{31}^{2}-f_{10}^{2} & f_{32}f_{31}-f_{20}f_{10} & 0\\
0 & f_{32}f_{31}-f_{20}f_{10} & 1+f_{32}^{2}-f_{20}^{2} & 0\\
f_{20}f_{23}+f_{13}f_{10} & 0 & 0 & 1+f_{13}^{2}+f_{23}^{2}%
\end{array}
\right)  . \tag{10}%
\end{equation}

Notice that we will not take into account general powers of the geometrical
Lagrangian (1), as for example $L_{g}=\left\vert \det G_{\mu\nu}\right\vert
^{\frac{u}{2}}$, because by taking the `square root form' ($u=1$) the
Lagrangian linearizes (as in the Yang-Mills case), and the solutions can be
easily compared to other cases in the literature \footnote{Geometrical
Lagrangians with generalized powers are sometimes convenient when the symmetry
of the considered coordinates coincides with the symmetry of the group
structure of the spacetime manifold, as in that cases $\left\vert \det
G_{\mu\nu}\right\vert $ becomes a perfect square and the Lagrangian is
automatically linearized (becomes of `Yang-Mills-type'), with the consequent
destruction of the nonlinear and nonlocal character of the solutions [6]}.

The\ physical magnetic field will be directly involved with the spin (physical
electric fields lead null matrix element contribution to the cross section) ,
consequently we must take $f_{13}=f_{23}=0.$ The resulting Lagrangian is%
\begin{equation}
\sqrt{\left\vert \det G_{\mu\nu}\right\vert }=\left(  2\lambda\right)
^{2}\left(  1-f_{20}^{2}-f_{10}^{2}\right)  \tag{11}%
\end{equation}
and%
\begin{equation}
\mathbb{F}_{20}=-2\left(  2\lambda\right)  ^{2}f_{20},\qquad\mathbb{F}%
_{10}=-2\left(  2\lambda\right)  ^{2}f_{10} \tag{12}%
\end{equation}
In this case the set of equations (8,9) reduces to%
\begin{align}
\partial_{r}\left(  f_{20}r\right)  -\partial_{\theta}f_{10}  &  =0\tag{13}\\
\partial_{\theta}\left(  \mathbb{F}_{20}r\sin\theta\right)  +\partial
_{r}\left(  \mathbb{F}_{10}r^{2}\sin\theta\right)   &  =0\nonumber
\end{align}
that leads immediately to the following solution for the physical magnetic
fields%
\begin{equation}
f_{10}=-\frac{\varphi_{23}}{j}=\frac{2\mu}{r^{3}}\cos\theta,\qquad
f_{20}=\frac{\varphi_{13}}{j}=\frac{\mu}{r^{3}}\sin\theta\tag{14}%
\end{equation}
and
\begin{equation}
\mathbb{F}_{10}=-\frac{\Phi_{23}}{j}=-2\left(  2\lambda\right)  ^{2}\frac
{2\mu}{r^{3}}\cos\theta,\qquad\mathbb{F}_{20}=\frac{\Phi_{13}}{j}=-2\left(
2\lambda\right)  ^{2}\frac{\mu}{r^{3}}\sin\theta\tag{15}%
\end{equation}
From the above equations the $h_{a}$ vector can be easily computed from the
dynamical equation [5,6,7]for the dual of the torsion field
\begin{equation}
dh=-\lambda^{\ast}f \tag{16}%
\end{equation}
We obtain
\begin{align}
&  h_{3}\omega^{3}=h_{\varphi}d\varphi\tag{17}\\
\text{with}  &  \text{ \ }h_{\varphi}=-\lambda\frac{\mu\sin^{2}\theta}{r}%
\quad\text{or well}\quad\rightarrow h_{3}=-\lambda\frac{\mu\sin\theta}{r^{2}}
\tag{18}%
\end{align}

\section{Cross-section}

\label{sec3}

We consider the spin flip of a neutrino coming in the $z$-direction over a
fixed source (\emph{e.g.} a hadron). We now assume, as usual: $\psi_{i}%
=N_{i}u\left(  k_{i},\uparrow\right)  e^{-ik_{i}x}$ and $\psi_{f}%
=N_{f}u\left(  k_{f},\downarrow\right)  e^{-ik_{f}x}$ then, the transition
amplitude is
\begin{equation}
S_{fi}=\delta_{fi}-\frac{i}{4}\left(  \frac{j\lambda}{d}\right)  \int
d^{4}x\overline{\psi}_{f}\gamma_{5}\gamma^{\mu}h_{\mu}\psi_{i} \tag{19}%
\end{equation}

The invariant amplitude is given by the matrix element%
\begin{equation}
\mathcal{M}=\frac{i}{4}j_{5}^{\mu}h_{\mu}\left(  q\right)  \tag{20}%
\end{equation}
where the axial current
\begin{equation}
j_{5}^{\mu}=\overline{u}_{f}\gamma_{5}\gamma^{\mu}u_{i} \tag{21}%
\end{equation}
in our case takes the form
\begin{equation}
\overline{u}_{f}\left(  p\uparrow\right)  \gamma_{5}\gamma^{3}u_{i}\left(
p\downarrow\right)  \tag{22}%
\end{equation}
We also have
\begin{equation}
h_{\mu}\left(  q\right)  =\int e^{iq\cdot x}h_{\mu}\left(  x\right)  d^{3}x
\tag{23}%
\end{equation}
Considering $h_{3}\left(  x\right)  =-\lambda\frac{\mu\sin\theta}{r^{2}}$ and
therefore using
\begin{equation}
\gamma^{5}\gamma^{3}h_{3}\left(  x\right)  =\gamma^{5}\left(  \gamma^{x}%
\sin\varphi+\gamma^{y}\cos\varphi\right)  (d-1)\frac{\mu}{r^{2}}, \tag{24}%
\end{equation}
where the geometrical relations $d\varphi=\frac{1}{x^{2}+y^{2}}\left(
xdy-ydx\right)  $ and $r\sin^{2}\theta d\varphi=\left(  \cos\varphi
dy-\sin\varphi dx\right)  $ were introduced in order to change the tetrad
basis to the coordinate one, the axial current in this case can be written
exactly as
\begin{equation}
\overline{u}_{f}\left(  p,\uparrow\right)  \gamma_{5}\gamma^{3}h_{3}%
(x)u_{i}\left(  p,\downarrow\right)  =\frac{p_{z}}{\left(  E+m\right)
}(d-1)\frac{\mu}{r^{2}}, \tag{25}%
\end{equation}
(notice that the Fourier transform was not performed yet). Then the matrix
element is immediately written as
\begin{equation}
\mathcal{M}=2\frac{(d-1)\mu}{\pi^{2}\left(  E+m\right)  }\ln\left\vert
\frac{q}{q_{\min}}\right\vert , \tag{26}%
\end{equation}
where $m$ is the neutrino mass. In the low momentum transfer or elastic
scattering limits (as in the computation of Ref. [2,16]) we obtain%
\begin{equation}
\mathcal{M}\simeq\frac{(d-1)\mu}{\pi^{2}\left(  E+m\right)  }\ln\left\vert
\frac{2\left(  E^{2}-m^{2}\right)  \left(  1-\cos\beta\right)  }{q_{\min}^{2}%
}\right\vert \tag{27}%
\end{equation}
with $\beta$ the scattering angle and the $q_{\min}$ can be associated to a
suitable cutoff $\Lambda$ (see below).

From [8] we know that the Dirac equation derived from the same geometry of the
spacetime in our unified model reads
\begin{equation}
\left[  i\gamma^{\alpha}\left(  \nabla_{\alpha}+j\frac{1-d}{d}\gamma
_{5}h_{\alpha}\right)  -\frac{mc}{\hbar}\right]  \psi=0 \tag{28}%
\end{equation}
where $j$ is a parameter of a geometrical origin (as we pointed out before)
and $d$ is the spacetime dimension. This formula (28) is, as expected,
completely analogous to other similar expressions on the literature coming
from non standard theories involving torsion with minimal coupling.

The expression for the differential cross section can be easily shown to be%
\begin{equation}
\frac{d\sigma}{d\Omega}=\left(  j\frac{1-d}{d}\right)  ^{2}\frac{E^{2}%
}{\left(  2\pi\hbar c\right)  ^{2}}\left\vert \mathcal{M}\right\vert ^{2}
\tag{29}%
\end{equation}
From this we obtain the important result%
\begin{equation}
\sigma_{\nu}^{flip}=\left(  \frac{j\mu mc}{4\hbar}\right)  ^{2}\left(
\frac{\left(  1-d\right)  ^{2}}{\pi^{2}d}\right)  ^{2}\frac{E^{2}}{\left(
E+mc^{2}\right)  ^{2}}\left[  \ln\left\vert \frac{q^{2}}{q_{\min}^{2}%
}\right\vert \right]  ^{2} \tag{30}%
\end{equation}
Notice the important fact that a similar expression was obtained in [2]for the
anomalous angular momentum (with anomalous momentum, in the notation of [2],
$\kappa\sim\left(  j\frac{1-d}{d}\right)  (d-1)\mu$).

In the elastic limit or the low momentum transfer we obtain
\begin{equation}
\sigma_{\nu}^{flip}\approx\left(  \frac{j\mu mc}{4\hbar}\right)  ^{2}\left(
\frac{\left(  1-d\right)  ^{2}}{\pi^{2}d}\right)  ^{2}\frac{E^{2}}{\left(
E+mc^{2}\right)  ^{2}}\left[  \ln\left\vert \frac{2\left(  E^{2}-m^{2}%
c^{4}\right)  \left(  1-\cos\beta\right)  }{q_{\min}^{2}}\right\vert \right]
^{2} \tag{31}%
\end{equation}
Notice the explicit dependence on the energy, in sharp contrast with the
string theoretical result [18] and the standard model one [16].

\section{Solar neutrino situation and spin-flip}

\label{sec4}

To compare our results with the ones coming from string or the standard model
theoretical considerations is useful to take the same notation and similar
approximations that ref.[16,18] for example. Then, in the case of elastic
scattering or low energy transfer, the probability of spin-flipping is given
by
\begin{equation}
\sigma_{\nu}^{flip}=A\frac{E^{2}}{\left(  E+mc^{2}\right)  ^{2}}\left[
\ln\left\vert B\left(  1-\cos\beta\right)  \right\vert \right]  ^{2} \tag{32}%
\end{equation}
where we have defined%
\begin{equation}
A\equiv\left(  \frac{j\mu mc}{4\hbar}\right)  ^{2}\left(  \frac{\left(
1-d\right)  ^{2}}{\pi^{2}d}\right)  ^{2},B\equiv\frac{2\left(  E^{2}%
-m^{2}c^{4}\right)  }{q_{\min}^{2}} \tag{33}%
\end{equation}
Notice that in strong coincidence with the results of several experimental
data [12,19,17,1], the cross section (32)\ is explicitly energy dependent, and
in our case we have, in addition, the geometrical parameter $j$ and the
dimension $d.$ In order to evaluate phenomenologically the problem, we must
integrate $\sigma$ around the scattering angle $\beta,$ as usual%
\begin{align}
\sigma_{\nu}^{flip}\left(  \beta\right)   &  =A\frac{E^{2}}{\left(
E+mc^{2}\right)  ^{2}}\overset{\equiv I}{\overbrace{\int\sin\beta
d\beta\left[  \ln\left\vert B\left(  1-\cos\beta\right)  \right\vert \right]
^{2}}}\tag{34}\\
\sigma_{\nu}^{flip}\left(  \beta\right)   &  =A\frac{E^{2}}{\left(
E+mc^{2}\right)  ^{2}}4\left[  2+\left(  Ln\left(  2\right)  \right)
^{2}-Ln4+LogB\left(  -2+Ln4+LogB\right)  \right]  \tag{35}%
\end{align}

where we have assigned the letter $I$ to the integral involving the scattering
angle $\beta$, in order to compare in a clear manner with the corresponding
string theoretical result or similar processes in the standard model. The
explicit form of the above formula for $\sigma_{\nu}^{flip}\left(
\beta\right)  $ is
\begin{align}
\sigma_{\nu}^{flip}\left(  \beta\right)   &  =\left(  \frac{j\mu mc}{4\hbar
}\right)  ^{2}\left(  \frac{\left(  1-d\right)  ^{2}}{\pi^{2}d}\right)
^{2}\frac{E^{2}}{\left(  E+mc^{2}\right)  ^{2}}\cdot\tag{36}\\
&  \cdot4\left[  1.09416+Ln\left(  \frac{2\left(  E^{2}-m^{2}c^{4}\right)
}{q_{\min}^{2}}\right)  \left(  Ln\left(  \frac{2\left(  E^{2}-m^{2}%
c^{4}\right)  }{q_{\min}^{2}}\right)  -0.613706\right)  \right] \nonumber
\end{align}

From the above results, it is important to note the following:

i) In formula (36), if we assume some astrophysical implications as in
Ref.[16], the logarithmic terms can be bounded with values between 1 and 6,
depending on screening arguments, as is generally accepted. For the
logarithmic terms close to 1, $I$ is approximately 6. On the other hand, the
string theoretical value of Ref. [18] takes the value $I=4.14$ (obtained
numerically). This situation of taking the logarithmic energy dependent terms
to be constant is at present questioned by the experimental point of view due
to the arguments given in the Introduction.

ii) The $j$ parameter plays formally (at the cross section level) a role
similar to that of the constant $\kappa$ of the string model with torsion of
[18]. However in our approach, it is related to some physical "absolute field"
(as $b$ in the Born-Infeld theory case, as we discuss in the Introduction)
giving the maximum value that the physical fields can take into the spacetime
(as the light velocity $c$ in the relativity theory). In such a case $j$ ("the
absolute field") will be fixed to some experimental or phenomenological value.

iii) The above results can be straighforwardly applied to the solar neutrino
case (e.g. a neutrino emerging from the Sun). If we call $P\left(  I\right)  $
the probability that the neutrino suffers an helicity change of spin and $n$
being the number of scattering centers per unit of volume, the probability of
a helicity flipping for the neutrino is%
\begin{equation}
P_{flip}=\int_{0}^{R_{\odot}}\sigma_{\nu}^{flip}ndl\approx\sigma_{\nu}%
^{flip}\overline{n}R_{\odot} \tag{37}%
\end{equation}
where, to get the estimate, we have taken the $n$ value in the average
$\overline{n}=10^{24}cm^{-1}$ and the Sun radius $R_{\odot}=10^{11}cm$ . This
expression has the geometrical parameter $j$ (see 36) as the link between the
experimental data and the theory, as we will see soon.

\section{Discussion and examples}

\label{sec5}

Now we will bring a few simple examples to discuss roughly the meaning of the
results given in the previous sections in the light of some experimental
results. At lower energies of the neutrino $<0.1MeV$, [15,22,4] the magnetic
momentum is around 10$^{-11}\mu_{B}$ \ but at higher energies, but with the
same bounded magnetic momentum of the neutrino, the cross section can be
written as a function of the cross section for the spin flipping due to the
Z$^{0}$ exchange in the weak interaction%
\begin{equation}
\frac{d\sigma}{d\Omega}=\left(  j\frac{1-d}{d}\right)  ^{2}\frac{2G_{F}%
^{2}m^{2}}{3\pi}\sin^{4}\theta_{W} \tag{38}%
\end{equation}

Notice that, if we consider the specific knowledge of the neutrino magnetic
momentum, and the experimental probability (cross section) of the flipping
then, the $j$ parameter (for instance $b$, the absolute field) will be
partially fixed.

\subsection{Comparison with string-gravity model with torsion potential}

In [18], the cross section for the minimal coupling condition in the gravity
string theory with torsion potential is%
\begin{equation}
\left.  \sigma_{\nu}^{flip}\left(  \beta\right)  \right\vert _{string}%
\approx8.28\pi\left(  \frac{9GSm}{4\hbar c^{2}}\right)  ^{2} \tag{39}%
\end{equation}
where $G$ is the Newton constant and $S$ is the spin of the particle.

In our theory, a rough comparison with the string model of [18] only can be
made assuming, as we have pointed out before, the logarithmic terms\ in
(36)close to 1 (or close to any other value between 1 and 6), and $d=4$:
\begin{equation}
\sigma_{\nu}^{flip}\left(  \beta\right)  \approx1.5\left(  j\frac{9}{\pi^{2}%
4}\right)  ^{2}\left.  \sigma_{\nu}^{flip}\left(  \beta\right)  \right\vert
_{string}. \tag{40}%
\end{equation}

precisely in order to avoid the lack of energy dependence at high energies of
the string theoretical result. Notice the existent relation between the $j$
parameter and the experimentally detected probability of flipping $P$. This
fact will be pointed out in the Concluding Remarks.

\subsection{Cross sections energy-dependent and logarithmic terms}

As mentioned previously, the neutrino counts have been experimentally obtained
for different energy ranges generating a clear preference on the cross
sections with energy windows as is the case of the Bethe computations[2] and
our own results computed here. This fact also evidently demands to review and
to correct the claim about the constancy of the logarithmic terms involving
the energy of the cross sections (as expression (36) or in the cross section
computation of [2]).

\section{Interaction structure, axions and Nambu-Goldstone bosons}

Theoretical arguments in a phenomenological context suggest that many
symmetries of the nature involving fundamental particles are spontaneously
broken. Having this fact into account, now we can show that a \textit{concrete
relation} between the axial vector $h_{\alpha}$ appearing of our model and the
axion field $a$ exists: the reason of our claim is as follows. We focusing now
in pseudoscalars: the first example are axions which from long time ago were
proposed as a possible solution to the (strong) CP problem [20,21]. Actually,
axions are only \textquotedblleft pseudo Nambu-Goldstone
bosons\textquotedblright\ in that the spontaneously broken chiral Peccei-Quinn
symmetry $U_{PQ}(1)$ is also explicitly broken, providing these particles with
a small mass
\[
m_{a}=0.60eV\frac{10^{7}GeV}{f_{a}}%
\]

where, $f_{a}$ is an energy scale (sometimes called of Peccei-Quinn). This
scale is related to the vev (vacuum expectation value) of the field which
breaks explicitly the Peccei-Quinn symmetry $U_{PQ}(1).$This scale is the main
quantity to be astrophysically constrained due that the Nambu-Goldstone
properties are directly related to it.. Notice that one can express limits on
$f_{a}$ in terms of $m_{a}$. from the above equation due that it is specific
to axions. To calculate the energy-loss rate of the axion fields from stellar
plasmas, the interaction with the medium constituents must be specified. In
general, the interaction with a fermion with mass $m_{f}$ is of 2 types, namely%

\[
L_{1}\approx\frac{C_{f}}{2f_{a}}\overline{\psi}_{f}\gamma^{\alpha}\gamma
_{5}\partial_{\alpha}a\psi_{f}%
\]
or%
\[
L_{2}\approx-i\frac{m_{f}C_{f}}{f_{a}}\overline{\psi}_{f}\gamma_{5}a\psi_{f}%
\]
where $\psi_{f}$ is the fermion and $a$the axion field, $C_{f}$ is a
model-dependent coefficient of order unity and $g_{af}\equiv\frac{m_{f}C_{f}%
}{2f_{a}}$ plays the role of a Yukawa coupling. If we compare the interaction
Lagrangians $L_{1}$ and $L_{2}$ and \ the interaction coming from the Dirac
equation (28) derived in our unified model:
\[
L_{int}\approx\overline{\psi}_{f}\frac{1-d}{d}j\gamma^{\alpha}\gamma
_{5}h_{\alpha}\psi_{f}%
\]
we can easily see that precisely \textit{only} the $L_{1}$ (derivative form)
is related with $L_{int}$ provided that
\[
\partial_{\alpha}a\equiv h_{\alpha}\text{ \ \ \ \ }and\text{ \ \ \ }%
\frac{C_{f}}{2f_{a}}\sim\frac{1-d}{d}j
\]
and L$_{2}$ is automatically ruled out. This fact is \textit{largely
consistent with the current research}: as is well known, the interaction
involving derivatives of the pseudoscalar is more fundamental in the sense
that it respects the Nambu-Goldstone nature of these particles: e.g.: is
invariant under $a\rightarrow a+a_{0}$. Contrarily, the pseudoscalar fashion
of $L_{2}$ is supposed to be equivalent in the \textit{usual sense}, because
there are many technical problems and troubles when one try to calculate
different processes where two Nambu-Goldstone bosons are attached to one
fermion line ( for example an axion and a pion attached to a nucleon). [11].
Then, our remark here is that there exist a closed relation between the affine
geometrical models with torsion as the analyzed here, where the lagrangian
coming from the breaking of some symmetries (Goldstone) and the
phenomenological observations concerning interaction with axions:. the
frequently expected interaction term of derivative type naturally arises from
our model.

\section{Neutrino oscillation and IceCube data: constrained the torsion effect
at higher energies}

As is well known, the phenomenon of oscillation gives a concrete and
acceptable description of atmospheric neutrinos. Recently, the analysis of
IceCube data\ [26]has provided the first significant detection (%
$>$
5$\sigma$ ) of atmospheric neutrino oscillations at energies near the 25 GeV
oscillation maximum for vertical events. The measured oscillation parameters
are in good agreement with results from other experiments that have measured
the atmospheric oscillation parameters with high resolution at lower energies,
some of them we have been mention in the Introduction. Then, the measurements
agree with the theoretical predictions of the standard three-neutrino flavor
oscillation framework that, in the context of the new physics effects (where
torsion effects are included), means that the flipping effect has little
significance at such energy window, as described by the IceCube data. However,
(see [26] ) it is interesting to note that at lower energies, non oscillation
effects seems to become a little bit more significant, given the possibility
to constrain the torsion contribution to the neutrino problem and obtain
concrete quantities in the cross section formula (36) for example. It is
precisely, the scope of our actual research where the torsion effects must be
compared and analyzed in the framework of the LHC also.

\section{Concluding remarks}

\label{sec6}

In this letter the effect of the torsion field on the spin interaction was
specifically analyzed in the case of a model corresponding to an unified field
theory based on a generalized affine geometry, metric and with totally
antisymmetric torsion field.. The main point to remark here is that, although
the oscillation mechanism explain in a great meaning the solar neutrino
problem, some little part of the trouble still remains and the discrepancy can
be explained by the non standard interaction with\ the torsion field arising
from this geometrical model. The main results of technical character are:

\begin{itemize}
\item[1)] The cross section is energy dependent, even at high energies. It
also depends on the spacetime dimension, and on a parameter $j$ of pure
geometrical origin, probably associated to a scale or limiting value for the
antisymmetric 2-form field in the geometrical Lagrangian.

\item[2)] This geometrical parameter plays a completely analogous role in our
theory to that of the absolute $b$ field of the Einstein-Born-Infeld theory.
The analogy appears because $j$ is associated to the 2-form $f$ (the potential
of the torsion field), homogenizing the units at the Lagrangian level and
putting some limiting value to $f$, exactly as in the Born-Infeld case.

\item[3)] Contrarily to the string theory case of .[18] where there exist some
$K$ \ that is a free parameter, the $j$ parameter is not completely free:
there are physical constraints and phenomenological estimations coming from
quantum and classical backgrounds that must indicate the specific freedom on
his parameter (e.g. some bound over $j$ can be obtained from classical
solutions plus Huges-Drever experiments)

\item[4)] The Lorentz symmetry and the minimal coupling are not violated due
to the group structure of this specific model. The weak symmetry, as was
explicitly shown here, is slightly violated.

\item[5)] Additionally, the equivalence principle, as was demonstrated in
several references [5,6,7,23,24,25], is not violated due to the totally
antisymmetric character of the torsion field and the metric character of our theory.

\item[6)] Is clear that the cross-section computed here can not be directly
compared with the cross section computed from the string gravity model with
torsion potential of [18]. This fact is because the cross-section of .[18] has
not energy dependent terms (even at high energies).

Will be very interesting to use the results from this new unified model in
order to see the effects of the torsion in the treatment of the anomalies.
This issue involving quantum field theoretical methods of non-local and not
perturbative character (instantons, etc), are now under advanced research.

\item[7)] There exist a closed relation between the affine geometrical models
with torsion and the phenomenological observations concerning interaction with
axions:. the frequently expected interaction term of derivative type naturally
arises from our model.

\item[8)] Because the new results from the IceCube data certainly ruled out at
higher energies non-standard interactions, in the case of the atmospheric
neutrinos the entire energy range must to be covered and analyzed to discern
the effects of torsion from other non-standard interactions. These effects
also must be constrained and compared with the possible torsion effects in the LHC\ framework.
\end{itemize}

Although the motivation of this work is clear we stress that the Standard
Model of particle physics (SM) successful theory till today, evidently need to
be revised due the observation of signs of a new physics beyond the scope of
the SM. Which kind of a new physics can be expected from the astrophysical
point of view? Evidently all aspects involving the inability of the SM to
incorporate quantum gravity.

\section{Acknowledgements}

Many thanks are given to Professor Yu. P. Stepanovsky for my scientific
formation and particularly to Dr. Victor I. Afonso for several discussions in
the subject and his help in the preparation of this text. I am very grateful
to Professor Xu Xin for introduce me into the affine geometrical models and
gravitation and also to Professors J.W.F. Valle and F.J. Escrihuela for bring
me very important references clearifying the subject of this paper and to
Professor George Raffelt for very useful insights and discussions.

\section{References}

[1] P. Anselmant \& al., 1993, GALLEX\ Colaboration, Phys. Lett B \textbf{314}.

[2] H.Bethe, 1935, Proc. Cam. Phyl. Soc, \textbf{31}, 108.

[3] S.Capozziello, G.Iovane, G.Lambiase \& C.Stornaiolo, 1999, Europhys.Lett.
\textbf{46}, 710-715.

[4] V. Castellani \& S. Degl'Innocenti, 1993, Astrophys. J. \textbf{402}, 574.

[5] D.J. Cirilo-Lombardo, 2010, Int.J.Theor.Phys. \textbf{49} 1288-1301.

[6] D.J. Cirilo-Lombardo, 2011a, Int.J.Theor.Phys. \textbf{50}, 1699-1708.

[7] D.J. Cirilo-Lombardo, 2011b, Int.J.Theor.Phys. \textbf{50}, 3621-3634.

[8] D.J Cirilo-Lombardo, Physics of Particles and Nuclei, 2013, Vol. 44, No.
5, pp. 848--865;

[9] D.J. Cirilo-Lombardo, 2007, J.Math.Phys.\textbf{\ 48} 032301

[10] D.J. Cirilo-Lombardo, 2005, Class.Quant.Grav. \textbf{22} 4987-5004.

[11] Choi K., Kang K., \& Kim J.E., 1989, Phys. Rev. Lett. 62, 849

[12] R Davis, 1992,\textit{\ Proc. Int. Symp. on Neutrino Astrophysics
(Takayama/Kamioka, 1992)}, ed . Y Suzuki and N. Nakamura (Tokio, Universal
Academy) p.47.

[13] F.J. Escrihuela, O.G. Miranda, M.A. Tortola \& J.W.F. Valle, 2009,
Phys.Rev. D80, 105009,

[14] F.J. Escrihuela, M. Tortola, J.W.F. Valle \& O.G. Miranda, 2011,
Phys.Rev. D83, 093002

[15] M. Fukugita \& S. Yazaki, 1987,Phys. Rev. D \textbf{36}, 3817.

[16] K. J. F. Gaemers \& al., 1989, Phys. Rev. D \textbf{40}, 309--314.

[17] V. N. Gavrin, 1993, \textit{Proc. TAUP\ Workshop (19-21 Sept.1993, Gran
Sasso Nat. Laboratory, Aquila, Italy)}.

[18] R. Hammond, 1996,Class.Quant.Grav. \textbf{13,} 1691-1697.

[19] K. S. Hirata \& al., 1991 Phys. Rev Lett \textbf{66}, 9.

[20] Peccei R.D.\& Quinn H.R., 1977a, Phys. Rev. Lett. 38, 1440

[21] Peccei R.D.\& Quinn H.R., 1977b,Phys. Rev. D 16, 1791;

[22] G. G. Raffelt, 1990, Astrophys. J. \textbf{365}, 559.

[23] Yu Xin, 1996, \textquotedblleft General Relativity on Spinor-Tensor
Manifold\textquotedblright, in: \textquotedblleft Quantum Gravity - Int.School
on Cosmology \& Gravitation\textquotedblright, XIV Course. Eds. P.G. Bergman,
V. de. Sabbata \& H.J. Treder, pp. 382-411, World Scientific.

[24] Yu Xin, 1989, Ap.SS \textbf{154}, 321.

[25] Yu Xin, 1993 Ap.SS \textbf{202}, 237.

[26] Measurement of Atmospheric Neutrino Oscillations with IceCube, IceCube
Collaboration, arXiv:1305.3909

\bigskip

\bigskip

\bigskip

\bigskip

\bigskip

\bigskip

\bigskip

\bigskip

\bigskip

\bigskip

\bigskip

\bigskip

\bigskip

\bigskip

\bigskip

\bigskip

\bigskip

\bigskip

\bigskip

\bigskip
\end{document}